\documentclass{PNASTWO}

\usepackage{cite}
\usepackage{graphicx}
\usepackage{color}
\usepackage{lmodern}

\usepackage{amssymb,amsfonts,amsmath}

\usepackage{PNASTWOF}

\newcommand{\rr}{{\bf{r}}}

\newcommand{\nn}{{\bf{n}}}
\newcommand{\mm}{{\bf{m}}}
\newcommand{\pp}{{\bf{p}}}
\newcommand{\CC}{{\bf{C}}}
\DeclareMathOperator{\trace}{tr}
\newcommand{\rref}[1]{{\bf \ref{#1}}}
\contributor{Published in Proceedings
of the National Academy of Sciences of the United States of America}
\url{www.pnas.org/cgi/doi/10.1073/pnas.1615729113}
\copyrightyear{2016}
\issuedate{November 15, 2016}
\volume{113}
\issuenumber{46}

\begin{document}


\title{A well-scaling natural orbital theory}





\author{Ralph Gebauer\affil{1}{ICTP -- The Abdus Salam International
  Centre for Theoretical Physics, Strada Costiera 11, 34151 Trieste,
  Italy}, Morrel H. Cohen\affil{2}{Department of Physics and
  Astronomy, Rutgers University, Piscataway, NJ
  08854}\affil{3}{Department of Chemistry, Princeton University,
  Princeton, NJ 08544}, \and Roberto
  Car\affil{3}{}\affil{4}{Department of Physics, Princeton University,
  Princeton, NJ 08544}}

\contributor{Published in Proceedings of the National Academy of Sciences
of the United States of America, vol. 113 no. 46 p. 12913 (2016)}

\significancetext{
Computations of locations of nuclei and movement of
electrons within molecules and materials are widely used in science
and technology. Direct computation of a system's wave function for
that purpose becomes impractical as system size grows. Current
alternative methods can have difficulty with strongly-correlated
electron motion or spurious electron self-interaction. By using
``natural spin orbitals'' to describe the motion of individual electrons,
solving for them together with their joint and individual probabilities
of occurrence within the system, we are able to account better for
electron correlation when strong while avoiding self-interaction and
maintaining the growth of computation cost with system size at the
level of Hartree-Fock theory. Our numerical results for some small
test molecules are very good. 
}

\maketitle

\begin{article}

\begin{abstract}
We introduce a new energy functional for ground-state electronic
structure calculations. Its variables are the natural spin-orbitals of
singlet many-body wave functions and their joint occupation
probabilities deriving from controlled approximations to the
two-particle density matrix that yield algebraic scaling in general,
and Hartree-Fock scaling in its seniority-zero version. Results from
the latter version for small molecular systems are compared with those
of highly accurate quantum-chemical computations. The energies lie above full
configuration interaction calculations, close to doubly occupied
configuration interaction calculations. Their accuracy is considerably
greater than that obtained from current density-functional theory
approximations and from current functionals of the one-particle
density-matrix. 
\end{abstract}






\section{Introduction}
\dropcap{C}omputing the ground-state energy of $N$ interacting electrons is
central to quantum chemistry, condensed-matter physics, and related
sciences. Reducing its complexity significantly below
that of the many-body wave function is a major
goal. Density-functional theory (DFT) \footnote{Section S8 of the
  online Supporting Information is a list of acronyms.}
\cite{Hohenberg1964, Kohn1965} achieved maximal reduction using
electron density as the basic variable. DFT transformed many sciences and
technologies, but finding accurate, parameter-free approximations to
its exchange-correlation energy functional that avoid self-interaction
and capture strong electron correlation is difficult.
One-particle density-matrix (1-DM) functional theories
\cite{Gilbert1975} have one
more degree of complexity. In them, the 1-DM is often
represented by its eigenvalues, the occupation numbers, and the
corresponding eigenvectors, the natural spin-orbitals (NSOs),
e.g.~\cite{Gritsenko2005, Lathiotakis2008}. While avoiding the
mean-field form of the 1-DM of DFT \cite{Kohn1965}, the 
approximations to the exchange-correlation functional of the 1-DM have
difficulties like those of the DFT approximations. Two-particle
density-matrix (2-DM) functional theories,
e.g.~\cite{Coleman1963a,Garrod1964,Percus2013}, are
less reduced. The ground-state energy is a
known, explicit functional of the 2-DM in Coulombic systems. However,
while useful complete conditions are known for the
$N$-representability of the 1-DM, the form of the known complete
conditions \cite{Garrod1964} for the 2-DM renders them unsuitable for
practical application. Nevertheless, major
progress has been 
made towards necessary conditions for $N$-representability that
can be systematically refined \cite{Zhao2004,Mazziotti2012}. While not
variational, the resulting calculations are almost as accurate as full
configuration interaction (FCI) calculations
\cite{Zhao2004,Nakata2001,Gidofalvi2008}. Their computational cost
scales as at least the 6th power of the basis-set size,
significantly worse than the asymptotic 3rd power scaling of
Hartree-Fock theory, DFT, and 1-DM theories.
In our new natural-orbital-functional theory, OP-NOFT, the basic
variables are the NSOs, their 
occupation numbers, and their joint occupation probabilities
(OP). That allows us to represent the 2-DM
accurately, transcending the  limitations of the 1-DM theories. Its
general form contains 
single-NSO through 4-NSO joint-occupation probabilities (n-OPs) and scales
as the 5th power of the basis-set size. Its simplest formulation, for
seniority 0, OP-NOFT-0, approximates doubly-occupied
configuration interaction (DOCI) \cite{Weinhold1967}. It contains only
1- and 2-natural-orbital (NO) OPs and retains the 3rd-power
scaling of Hartree-Fock energy-functional minimization with a higher prefactor. 
It describes the dissociation of simple diatomic molecules and
multi-atom chains with accuracy that can be comparable to that of
DOCI, which uses a compact basis of Slater determinants (SD) but retains
combinatorial scaling. 
In 4-electron systems, illustrated with the Paldus H$_4$ test
\cite{Jankowski1980}, it yields results identical to those of DOCI. 
OP-NOFT-0 is powerful at high correlation,
i.e.~for static correlation at intermediate and large interatomic
separations where Hartree-Fock fails due to the multi-reference character of the
ground-state wavefunction. There, OP-NOFT-0 outperforms Hartree-Fock,
DFT with standard approximations, and
quantum-chemistry methods such as (single-reference) coupled cluster
with single, double and perturbative triple electron-hole excitations
(CCSD(T)), a standard of accuracy near equilibrium separations. This
introduction of higher-order OPs as variational parameters, with
closure of the theory at their level, is the essential novelty of our
work and is responsible for its favorable scaling.

\section{OP-NOFT, general formulation}
\subsection{The NSO basis}

We consider time-reversal invariant saturated systems
with non-degenerate, singlet ground states. The inverse approach
\cite{Coleman1963a,Garrod1964,Percus2013} starts from the
$N$-representability conditions 
on the 2-DM. Instead, we take a forward approach: we
introduce a specific form for the trial wavefunction and derive
the 2-DM explicitly. Our starting point
is that of conventional FCI, except that our
one-particle basis is the complete set of NSOs 
of the trial function $\Psi$,
$\psi_k(x) = \phi_k(\rr) \chi_k(\sigma)$, 
with $\rr$ space and $\sigma$ spin
coordinates. The NOs $\phi_k(\rr)$ can be real and are 
independent of the spin function. The
complete set of $N$-electron 
orthonormal SDs $\Phi_{\nn}(x_1,x_2,\cdots,x_N)$, 
$\nn=k_1,k_2,\cdots,k_N$, formed from its NSOs 
supports representation of any trial wavefunction
$\Psi(x_1,x_2,\cdots,x_N)$ as the expansion
\begin{equation}
\Psi(x_1,\cdots,x_N) = \sum_{\nn} C_{\nn} \Phi_{\nn}(x_1,\cdots,x_N).
\label{expansion}
\end{equation}

As the ground-state wave function can be chosen to be real, so can be the
trial functions 
and the normalized $C_{\nn}$  ($\sum_{\nn} C_{\nn}^2 = 1$).
The NSOs vary with the trial function or the coefficients in the
search for the ground state.
The combinatorial
complexity of determining the ground-state energy by variation of the
$C_{\nn}$ is composed of the separate combinatorial complexities of the
signs and magnitudes of the coefficients $C_{\nn}$. We use
distinct reductive approximations for their signs and 
magnitudes. The signs 
depend on the sign convention chosen for the SDs. We use the Leibniz
form for the SDs,
\begin{eqnarray}
\Phi_{\nn}(x_1,\cdots,x_N) = \qquad \qquad \qquad\nonumber \\ 
\frac1{\sqrt{N!}} \sum_p \text{sgn}\{P_p\} P_p \psi_{k_1}(x_1) \cdots
\psi_{k_N}(x_N).
\label{SDs}
\end{eqnarray}
The sum is over the elements of the symmetric group of order $N$, the
permutations $P_p$. The sign of $\Phi_{\nn}$ is fixed by the ordering
$k_1 < k_2 < \cdots < k_N$ in the product of the NSOs $\psi_{k_i}$ in
\rref{SDs}. The SDs and their
coefficients can then be specified by listing the NSOs occupied
in the SDs, i.e.~by the index $\nn$.

\subsection{The 1-DM, the orthogonality constraint, the PDC}

The 1-DM of $\Psi$,
\begin{eqnarray}
\rho(x',x) = N \int dx_2\cdots dx_N \nonumber \\
\Psi(x',x_2,\cdots,x_N) \Psi(x,x_2,\cdots,x_N),
\label{rho}
\end{eqnarray}
becomes
\begin{equation}
\rho(x',x) = \sum_{i\neq j} \left[ \sum_{\mm \not\ni i,j} C_{i,\mm} C_{j,\mm}
  \right] \, \psi_i(x') \psi_j(x)
\label{rho1_spin}
\end{equation}
after \rref{expansion} and \rref{SDs} are inserted into
\rref{rho}. In \rref{rho1_spin} the subindex $\mm$ specifies 
the $N-1$ NSOs present in $\Phi_{i,\mm}$ and $\Phi_{j,\mm}$, 
excluding $\psi_i$ and $\psi_j$. As the $\psi$ are
the NSOs of $\Psi$, the eigenfunctions of $\rho(x',x)$,
the bracketed quantity in \rref{rho1_spin} must vanish for $i\neq
j$. Regard the coefficients $C_{i,\mm}$ and $C_{j,\mm}$
as the components of vectors $\CC_i$ and $\CC_j$
and the bracket as their scalar product $\CC_i \cdot \CC_j$,
which must vanish. There are two realizations of this
orthogonality constraint. In the first, and most general
form (OC), the presence of $\Phi_{i,\mm}$ in $\Psi$ does not exclude
the presence of $\Phi_{j,\mm}$. The individual terms in the
scalar product need not vanish. The second, the pair-difference
constraint (PDC), is 
a special case of the OC, in which the presence of $\Phi_{i,\mm}$
excludes $\Phi_{j,\mm}$ so that either 
$C_{i,\mm}$ or $C_{j,\mm}$ is zero for each $\mm$, and the sum
vanishes term by term. Under the PDC, those $\Phi$ present
in the expansion of $\Psi$ must differ from one another by at least
two NSOs. The OC is necessary and sufficient for $N$-representability, whereas
the PDC is only sufficient. We impose the PDC on the $\Phi$ as a
simplifying variational approximation. 
The PDC proved well satisfied in the FCI result for H$_8$
using the minimal basis set STO-6G.

Under the OC or PDC, $\rho(x',x)$ takes the diagonal
form 
\begin{equation}
\rho(x',x) = \sum_k p_1(k) \, \psi_k(x')\psi_k(x).
\label{rho1diag}
\end{equation}
Here, the $p_1(k) = \sum_{\nn} C_{\nn}^2 \, \nu_{k,\nn}$, where $\nu_{k,\nn}
= 1$ if $k \in \nn$ and $0$ otherwise, are the eigenvalues of
$\rho(x',x)$, the occupation numbers or occupation probabilities
(1-OP) of its eigenfunctions $\psi_k$. They satisfy the necessary and
sufficient conditions $0 \le p_1(k) \le 1$ and $\sum_k p_1(k) = N$. In
general, only $M > N$ occupation numbers $p_1(k)$ are non-negligible,
and only the corresponding {\em active} NSOs need be included
in the representation of any trial function, providing a natural
cutoff. The 1-DM is thus of algebraic complexity in the
1-OPs and the NSOs.

\subsection{The 2-DM, the sign conjecture, the $\xi$-approximation}

The 2-DM of $\Psi$,
\begin{eqnarray*}
\pi(x'_1x'_2;x_1x_2) = N (N-1) \int dx_3\cdots dx_N \\
\Psi(x'_1,x'_2,x_3,\cdots,x_N) \Psi^*(x_1,x_2,x_3,\cdots,x_N),
\end{eqnarray*}
becomes
\begin{eqnarray}
\pi(x'_1x'_2;x_1x_2) &=& 
\sum_{\scriptscriptstyle
\overset{i<i', j<j',\mm}
{i,i',j,j' \not\in \mm}}
C_{ii'\mm} C_{jj'\mm} \nonumber\\
&&\left(
\psi_i(x'_1)\psi_{i'}(x'_2)-\psi_{i'}(x'_1)\psi_i(x'_2)\right) \nonumber\\
&&\left(
\psi_j(x_1)\psi_{j'}(x_2)-\psi_{j'}(x_1)\psi_j(x_2)\right).
\end{eqnarray}
$\pi$ separates into a part $\pi^d$ diagonal in the indices, i.e.~with
$ii' = jj'$, and an off-diagonal part, 
$\pi^{od}$, with $ii' \neq jj'$:
\begin{eqnarray}
\pi^d(x'_1x'_2;x_1x_2) &&= \frac1{2}\sum_{i\ne j} p_{11}(ij) \nonumber\\
&&\left(
\psi_i(x'_1)\psi_j(x'_2)-\psi_j(x'_1)\psi_i(x'_2)\right)\nonumber\\
&&\left(
\psi_i(x_1)\psi_j(x_2)-\psi_j(x_1)\psi_i(x_2)\right)
\label{pi-d}\\
\pi^{od}(x'_1x'_2;x_1x_2) &&= 
\sum_{\scriptscriptstyle
\overset{i<i' \ne j<j',\mm}
{i,i',j,j' \not\in \mm}}
C_{ii'\mm} C_{jj'\mm}\nonumber\\
&&\left(
\psi_i(x'_1)\psi_{i'}(x'_2)-\psi_{i'}(x'_1)\psi_i(x'_2)\right)\nonumber\\
&&\left(
\psi_j(x_1)\psi_{j'}(x_2)-\psi_{j'}(x_1)\psi_j(x_2)\right)
\label{pi-od}
\end{eqnarray}

Electron correlation is expressed through
$\pi^{od}$. The analogous off-diagonal part of
$\rho(x',x)$ is suppressed by the OC, an advantage of the NSO
basis. Note that the PDC has eliminated 3-index terms from
$\pi^{od}$ in~\rref{pi-od}. 

The $p_{11}(ij) = \sum_{\nn} C_{\nn}^2 \, \nu_{i,\nn} \nu_{j,\nn}$ in
$\pi^d$ are joint 2-state occupation probabilities (2-OPs).
Mazziotti has reported \cite{Mazziotti2012, Mazziotti2012a} necessary and sufficient conditions on the
2-OPs that arise from the positivity conditions on the q-OPs, i.e.~the
$p_{11\cdots 1} (i_1,i_2,\cdots,i_q) = \sum_{\nn} C_{\nn}^2 \, \nu_{i_1,\nn}
\nu_{i_2,\nn} \cdots \nu_{i_q,\nn}$, at any order $2\le q\le N$. 
These conditions derive from the positivity
conditions \cite{Mazziotti2012a} on the diagonal elements of
the 2-DM \cite{Ayers2007}.
Limiting ourselves to the $(2,2)$ and $(2,3)$
conditions, the following conditions for the 2-OPs hold:
\begin{eqnarray}
\mbox{sup}(p_1(i)+p_1(j)-1,0) \le p_{11}(ij) \,\le\, p_1(<) \label{bound1}\\
\mbox{sup}(p_1(i)+p_1(j)+p_1(k)-1,0) \le
p_{11}(ij)+\nonumber\\
p_{11}(ik)+p_{11}(jk) \label{bound2}
\end{eqnarray}
$p_1(<)$ is the lesser of $p_1(i)$ and $p_1(j)$. In addition the sum rule
\begin{eqnarray}
\sum_{j (\ne i)} p_{11}(ij) &=& (N-1) p_1(i) \label{p11sum}
\end{eqnarray}
must be satisfied. Conditions \rref{bound1}--\rref{p11sum} were first
established by Weinhold and Bright Wilson \cite{Weinhold1967a}. 
They are necessary but not sufficient conditions for
$N$-representability
\cite{Ayers2007,Davidson1969,Garrod1964,Mazziotti2012}. 
Establishing a complete set of conditions is QMA-hard in $N$ because
the number of 
$(2,q)$ positivity conditions increases combinatorially with
increasing $q \le N$. 
Fortunately numerical calculations on atoms and molecules indicate
that sufficiently accurate lower-bound ground-state energies often
result by imposing $(2,q)$-positivity conditions with $q\le 3$
\cite{Zhao2004,Mazziotti2012b}. This
suggests that even in the most difficult situations, fermionic problems
in atoms and molecules should require only a finite and small set of positivity
conditions. Here we shall limit ourselves to conditions
\rref{bound1}--\rref{p11sum}, as we found in our numerical
calculations that they are sufficient to produce accurate
lower-bounds. If higher-order conditions were found to be necessary,
it would not be hard for us to add a few more.

The $\pi^d$ of \rref{pi-d} contains only 2-OPs and products of 2
distinct NSOs; it has at most algebraic complexity $\sim M^3$ deriving
from condition \rref{bound2}. Thus when only conditions
\rref{bound1}--\rref{p11sum} are imposed, the combinatorial
complexity of the ground-state problem resides entirely in the
$\pi^{od}$ of \rref{pi-od}. We extract
the sign $s(ii'\mm)$ of the coefficient $C_{ii'\mm}$ in
\rref{pi-od} and, relating its magnitude to the joint $N$-OP
$p_{11\cdots 1}(ii'\mm) \equiv C_{ii'\mm}^2$,
we rewrite~\rref{pi-od} as
\begin{align}
\pi^{od}(x'_1x'_2;x_1x_2) =& 
\sum_{\scriptscriptstyle
\overset{i<i'\ne j<j',\mm}
{i,i',j,j' \not\in \mm}}
s(ii'\mm) s(jj'\mm) \nonumber \\
&p^{1/2}_{11\cdots 1}(ii'\mm) 
p^{1/2}_{11\cdots 1}(jj'\mm)\nonumber\\
&\left(
\psi_i(x'_1)\psi_{i'}(x'_2)-\psi_{i'}(x'_1)\psi_i(x'_2)\right)\nonumber\\
&\left(
\psi_j(x_1)\psi_{j'}(x_2)-\psi_{j'}(x_1)\psi_j(x_2)\right).
\label{pi-od-signs}
\end{align}
We suppose that a variational approximation exists in which
\begin{equation}
s(ii'\mm) s(jj'\mm) = s(ii') s(jj'), \forall\mm.
\label{signconjecture}
\end{equation}
This {\em sign conjecture} reduces the sign
complexity to algebraic, scaling as $M^2$. $\pi^{od}$ 
simplifies to
\begin{eqnarray}
\pi^{od}(x'_1x'_2;x_1x_2) &=& 
\sum_{\scriptscriptstyle
{i<i' \ne j<j'}}
s(ii') s(jj') \nonumber \\
&&\left[ \sum_{\scriptscriptstyle \overset{\mm}
{i,i',j,j' \not\in \mm}}
p^{1/2}_{11\cdots 1}(ii'\mm) 
p^{1/2}_{11\cdots 1}(jj'\mm) \right] \nonumber\\
&&\left(
\psi_i(x'_1)\psi_{i'}(x'_2)-\psi_{i'}(x'_1)\psi_i(x'_2)\right)\nonumber\\
&&\left(
\psi_j(x_1)\psi_{j'}(x_2)-\psi_{j'}(x_1)\psi_j(x_2)\right).
\label{pi-od-signs1}
\end{eqnarray}
The quantities $p^{1/2}_{11\cdots 1}(ii'\mm)$ and $p^{1/2}_{11\cdots
  1}(jj'\mm)$ are $\mm$-th components of vectors
$\pp^{1/2}_{11\cdots 1}(ii')$ and $\pp^{1/2}_{11\cdots 1}(jj')$. The
bracketed quantity in~\rref{pi-od-signs1} is their scalar
product. Express it as
\begin{eqnarray}
\sum_{\scriptscriptstyle \overset{\mm}
{i,i',j,j' \not\in \mm}}
p^{1/2}_{11\cdots 1}(ii'\mm) 
p^{1/2}_{11\cdots 1}(jj'\mm) &=& \nonumber \\
p^{1/2}_{1100}(ii'jj') p^{1/2}_{0011}(ii'jj') \xi(ii'jj'),
\label{scalar}
\end{eqnarray}
where $p_{1100}(ii'jj')$ is the square magnitude of the vector
$\pp^{1/2}_{11\cdots 1}(ii')$ and $p_{0011}(ii'jj')$ that of
$\pp^{1/2}_{11\cdots 1}(jj')$. $p_{1100}(ii'jj')$ is the probability
that $\psi_i$ and $\psi_{i'}$ are occupied while $\psi_j$ and
$\psi_{j'}$ are not:
\begin{eqnarray*}
p_{1100}(ii'jj') &=& \sum_{\scriptscriptstyle \overset{\mm}
{i,i',j,j' \not\in \mm}} p_{11\cdots 1}(ii'\mm) \\
&=& \sum_{\nn} C_{\nn}^2 \, \nu_{i,\nn} \nu_{i',\nn} \, 
\left( 1 - \nu_{j,\nn} \right) \left( 1 - \nu_{j',\nn} \right),
\end{eqnarray*}
and the reverse is true for $p_{0011}(ii'jj')$. 

The Schwarz inequality $0 \le \xi(ii'jj') \le 1$ imposes bounds on
$\xi(ii'jj')$, the cosine of the hyper-angle between the
vectors. The upper bound $\xi =1$ is exact for
$N=2$. Substituting~\rref{scalar} into~\rref{pi-od-signs1} yields
\begin{eqnarray}
\pi^{od}(x'_1x'_2;x_1x_2) &=& 
\sum_{\scriptscriptstyle
i<i' \ne j<j'}
s(ii') s(jj') \nonumber\\
&&\left[p_{1100}(ii'jj') \,p_{0011}(ii'jj')\right]^{1/2}
\xi(ii'jj') \nonumber\\
&&\left(
\psi_i(x'_1)\psi_{i'}(x'_2)-\psi_{i'}(x'_1)\psi_i(x'_2)\right)\nonumber\\
&&\left(
\psi_j(x_1)\psi_{j'}(x_2)-\psi_{j'}(x_1)\psi_j(x_2)\right),
\label{final_pi-od}
\end{eqnarray}
in which only $\xi(ii'jj')$ retains combinatorial complexity:
\begin{displaymath}
\xi(ii'jj') = \frac{
\sum' p^{1/2}_{11\cdots 1}(ii'\mm) p^{1/2}_{11\cdots 1}(jj'\mm)}
{\left(
\sum' p_{11\cdots 1}(ii'\mm) \, \,
\sum' p_{11\cdots 1}(jj'\mm) 
\right)^{1/2}},
\end{displaymath}
where the primed sums are over all $\mm$ with $i,i',j,j' \not\in \mm$.

Inserting 4-OPs like
\begin{displaymath}
p_{1111}(ii'kl) = \sum_{\nn} C_{\nn}^2 \, \nu_{ii',\nn} \,
\nu_{kl,\nn}; \qquad k < l \neq i,i',j,j'
\end{displaymath}
in place of the $N$-OPs in $\xi$ reduces the complexity of $\pi^{od}$ to
algebraic. The resulting approximation,
\begin{equation}
\xi(ii'jj') \approx \frac{
\sum''_{k<l} p^{1/2}_{1111}(ii'kl) p^{1/2}_{1111}(jj'kl)}
{\left(
\sum''_{k<l} p_{1111}(ii'kl) \, \,
\sum''_{k<l} p_{1111}(jj'kl) 
\right)^{1/2}},
\label{xi-4}
\end{equation}
is not variational, but obeys the 0,1 bounds of the Schwarz
inequality. It is exact for $N=4$, and scales as
$M^4$. In~\rref{xi-4} the doubly-primed sums are over the indices
$k<l$, which must differ from $i,i',j,j'$.
Bounds on the $p_{1111}$ that are the
generalizations of \rref{bound1}--\rref{p11sum} for 3-OPs and 4-OPs
can be formulated. 

\subsection{The OP-NOFT energy functional}
The trial energy $E[\Psi] = \langle \Psi | \hat{H} | \Psi \rangle$, the
expectation value of the Hamiltonian $\hat{H}$, is an
explicit functional of the 1- and 2-DM:
\begin{displaymath}
E[\Psi] = E[\rho, \pi] = \trace\left\{\rho \hat{h}\right\} +
\trace\left\{ \pi \hat{w} \right\}.
\end{displaymath}
Here $\hat{h}$ is the single-particle kinetic-energy operator plus the
external potential, and $\hat{w}$ is the 2-electron Coulomb
interaction. $E[\rho, \pi]$ splits into two parts, $E^d$
diagonal and $E^{od}$ off-diagonal in the SD:
\begin{eqnarray}
E &=& E^d + E^{od} \nonumber \\
E^d &=& \trace\left\{\rho \hat{h}\right\} + \trace\left\{\pi^{d}
\hat{w}\right\} \nonumber \\
E^{od} &=& \trace\left\{\pi^{od} \hat{w}\right\}.
\label{Esum}
\end{eqnarray}
The Hartree-Fock wave function minimizes $E^d$; $\pi^{od}$ introduces electron
correlation into $E^{od}$. The explicit forms of $E^d$ and $E^{od}$
follow from \rref{rho1diag}, \rref{pi-d}, and \rref{final_pi-od}:
\begin{equation}
E^d = \sum_i p_1(i) h_{ii} + \sum_{i<j} p_{11}(ij) \left[ {\cal
    J}_{ij} - {\cal K}_{ij}\right],
\label{E-d}
\end{equation}
where $h_{ii} = \langle\psi_i | \hat{h} | \psi_i\rangle$, and  ${\cal J}_{ij} = 
\langle \psi_i\psi_i | \hat{w} | \psi_j\psi_j\rangle$ and ${\cal K}_{ij} = 
\langle \psi_i\psi_j |\hat{w}|\psi_j\psi_i\rangle$ are the Coulomb and exchange 
integrals, respectively.
The second
term on the rhs of \rref{E-d} originates from
$\pi^d$, the form of which is represented exactly in our theory. It contains only
positive contributions and is essential; the integral
relation connecting $\pi$ and $\rho$ depends only on $\pi^d$ and
guarantees that the $E$ is self-interaction free.
\begin{eqnarray}
E^{od} &=& \sum_{\scriptscriptstyle i<i' \ne j<j'} s(ii') s(jj')
p^{1/2}_{1100}(ii'jj') p^{1/2}_{0011}(ii'jj')
\nonumber \\
&& \xi(ii'jj') \left[ {\cal K}_{ii'jj'} - {\cal K}_{ii'j'j} \right],
\label{E-od}
\end{eqnarray}
where ${\cal K}_{ii',jj'} = \langle \psi_i\psi_{i'} | \hat{w} | \psi_j\psi_{j'}\rangle$.
\rref{Esum} -- \rref{E-od} define the OP-NOFT energy
functional within the PDC. Including the complexity of efficient
evaluation of the matrix elements, it scales as $M^5$ if the
$N$-representability conditions for the 3- and 4-OPs can be limited to
those deriving from the $(3,q)$ and $(4,q)$ positivity conditions with $q
\le 4$.  

\subsection{Proof of the sign conjecture}

A variational sign approximation must be a statement about the sign
$s(ii'\mm)$ or $s(jj'\mm)$  of each coefficient appearing
in~\rref{pi-od}. To prove \rref{signconjecture}, we must find at least one
statement in which the $\mm$-dependences of $s(ii'\mm)$ and $s(jj'\mm)$ 
cancel. We have found two and present one here and one in 
{section S1} of the SI. The former is valid for the
general case of matrix elements 
$\left[ {\cal K}_{ii',jj'} - {\cal K}_{ii',j'j} \right]$ of
arbitrary sign, the latter only for positive ones. 

Assigning each index $l$ in $C_{\nn}$ a sign $s(l)$ and taking
$s(\nn)$ as their product to form
$s(\nn)=\prod_{l\in \nn} s(l)$ is a variational approximation. 
Consequently $s(ii'\mm) = s(i)s(i')s(\mm)$, and
\begin{equation}
s(ii'\mm) \, s(jj'\mm) = s(i) s(i') s(j) s(j')
\label{signproduct}
\end{equation}
so that \rref{signconjecture} is proved, with $s(ii')=s(i)s(i')$.

This approximation treats the form and phase, 0 or $\pi$, of each NSO as
independent variables. The choice of signs for each index is not
specified in \rref{signproduct}. 
Most energy minimization schemes start with random initial NSOs; similarly the
choice of signs in \rref{signproduct} should be random, half positive
and half negative. The number of the initial NSOs should be greater
than the anticipated value of $M$ to allow for unequal numbers of
positive and negative signs of the active NSOs.

This complete factorization of $s(\nn)$ and thence of $s(ii')$ is a
restrictive approximation. A variational approximation
yielding unfactorized $s(ii')$ could be more accurate. In
{Section S1} of 
the SI we have introduced a different variational approximation and
rule for the signs which leads to \rref{signconjecture} without
factorization for positive matrix elements. 
Random assignment of signs in 
\rref{signconjecture} and the sign rule of 
{Section S1} yield identical results where tested, the significance of
which is discussed there.

\section{OP-NOFT-0}

The SD's in \rref{expansion} can be classified by their seniority,
the number $A$ of singly-occupied one-particle states they
contain. For $N$ even and for a global
spin singlet ($S=0$) state, the $N$-particle Hilbert space divides into
sectors of increasing even seniority starting with $A=0$, where all
SD's contain only doubly occupied states. 
For molecular systems CI expansions converge rapidly
with seniority, and DOCI $A=0$ calculations describe
static correlation rather well, as demonstrated in~\cite{Bytautas2011}. 

The PDC is equivalent to a restriction on seniorities in
that seniorities differing only by 4 are allowed. 
Recent CI calculations for systems with even numbers of electrons
showed that the seniority 2 sector largely 
decouples from the seniority 0 sector, supporting the accuracy of the
PDC \cite{Bytautas2011}. These considerations also apply to systems with an odd number of
electrons, in which seniorities would be odd but still differ only by
4. 

We now formulate OP-NOFT explicitly in the $A=0$ sector to
illustrate further how an OP-NOFT functional is constructed and to prepare  
for numerical implementation; it becomes OP-NOFT-0, in which the PDC
is automatically satisfied. Tracing out the spins,
\rref{rho1diag} becomes:
\begin{equation}
\rho(\rr',\rr) = 2 \sum_k p_1(k) \, \phi_k(\rr') \phi_k(\rr).
\label{rho1}
\end{equation}
$k$ now labels $M(>N/2)$ active doubly-occupied NO states, and the following conditions
hold:
\begin{equation}
0 \le p_1(k) \le 1 \mbox{ and } 2\sum_k p_1(k) = N.
\label{rho1cond}
\end{equation}
In \rref{rho1} and \rref{rho1cond} $p_1(k)$ is the occupation number of either 
of the paired NSOs having the NO $\phi_k$.

In the 2-DM, double occupancy results in a major simplification of the structure of 
$\pi^{od}$. We make the orbital and spin components of the NSO indices
explicit. They take the form $is$, with $i$ now the orbital index and $s = \pm$ the
spin index. The only index pairs that can enter
$\pi^{od}$ in \rref{final_pi-od} are $i+,i-$ and $j+,j-$. The
only sets of two index pairs that can enter the rhs of \rref{xi-4} are $i+,i-$,
$k+,k-$ and $j+,j-$, $k+,k-$. The occupation numbers $\nu_{i+}$ and 
$\nu_{i-}$ are equal, with values 0 or 1, so that all 4-NSO OPs in
\rref{xi-4} and \rref{final_pi-od} are identical to the corresponding 
spin independent 2-NO OPs, e.g.~$p_{1111}(i+,i-,k+,k-) = p_{11}(ik)$.
The signs in \rref{final_pi-od} depend on a single
orbital index, $s(i+,i-) = {\mathsf s}(i)$, and the $\xi$ depend on two-orbital
indices, $\xi(i+,i-,j+,j-) = \xi(ij)$. With these simplifications, the 2-DM of
\rref{pi-d} and \rref{final_pi-od} becomes
\begin{equation}
\pi(\rr'_1 \rr'_2;\rr_1 \rr_2) =
\pi^d(\rr'_1 \rr'_2;\rr_1 \rr_2) +
\pi^{od}(\rr'_1 \rr'_2;\rr_1 \rr_2),
\label{pisum}
\end{equation}
 after tracing out the spins, with
\begin{eqnarray}
\pi^d(\rr'_1 \rr'_2;\rr_1 \rr_2) &=& 
2 \sum_{ij} p_{11}(ij) \nonumber\\
&&\big(2 \phi_i(\rr'_1) \phi_j(\rr'_2) \phi_i(\rr_1) \phi_j(\rr_2) - \nonumber \\
&&\phi_i(\rr'_1) \phi_j(\rr'_2) \phi_j(\rr_1) \phi_i(\rr_2) \big) 
\label{pi-d_DOCI}\\
\pi^{od}(\rr'_1 \rr'_2;\rr_1 \rr_2) &=& 2 \sum_{i\ne j}
{\mathsf s}(i) {\mathsf s}(j) \nonumber \\
&&\left[ p_{10}(ij) p_{01}(ij)\right]^{1/2} \xi(ij) \nonumber \\
&& \phi_i(\rr'_1) \phi_i(\rr'_2) \phi_j(\rr_1) \phi_j(\rr_2).
\label{pi-od_DOCI}
\end{eqnarray}
The sum in \rref{pi-d_DOCI} includes the term $i=j$, for which $p_{11}(ii) =
p_1(i)$, and $\xi(ij)$ in \rref{pi-od_DOCI} is now
\begin{equation}
\xi(ij) \approx \frac{
\sum_{k(\ne i,j)} p^{1/2}_{11}(ik) p^{1/2}_{11}(jk)
}{\left[
\sum_{k(\ne i,j)} p_{11}(ik) \, \, \sum_{k(\ne i,j)} p_{11}(jk)
\right]^{1/2}}.
\label{xi_DOCI}
\end{equation}
The one- and two-orbital OPs of OP-NOFT-0 lie within the same bounds as 
in the general case, \rref{bound1}--\rref{bound2}, 
and their sum rules become, respectively, \rref{rho1cond} and
\begin{equation}
2 \sum_{j(\ne i)} p_{11}(ij) = (N-2) p_1(i).
\label{p11sum_DOCI}
\end{equation}
The $\pi$ of \rref{pisum} satisfies two important sum rules
\begin{eqnarray*}
\int d\rr_2 \, \, \pi(\rr \rr_2; \rr' \rr_2) &=& (N-1) \rho(\rr,\rr') \\
\int d\rr_1 \, d\rr_2 \, \, \pi(\rr_1 \rr_2;\rr_1 \rr_2) w(r_{12})
&\ge& 0. 
\end{eqnarray*}

The OP-NOFT-0 form for $\pi$, \rref{pisum}--\rref{xi_DOCI}, 
is exact 
when $N=2$ with $\xi=1$. It is equivalent to DOCI for $N=4$ if the
signs are correct. 
We show numerically that this is the case for the sign choice
of {Section S1} for the Paldus H$_4$ test, 
as reported in {Section S5} of the SI.
For all the H$_4$ configurations studied, the OP-NOFT-0 correlation
energy coincides with that of DOCI to numerical precision. That the
signs are correct for H$_2$ and H$_4$ confirms the validity of the
sign rule in those cases and suggests a broader utility.
When $N>4$, the $\xi$-approximation of \rref{xi_DOCI}
and the limitation to the (2,2) and (2,3) positivity conditions
break the equivalence to DOCI. 

The expectation value $E = \langle \Psi | \hat{H} |
\Psi \rangle$ becomes: 
\begin{eqnarray}
E &=& 2 \sum_i p_1(i) \langle\phi_i | \hat{h} | \phi_i\rangle + \nonumber \\
&&\sum_{ij} p_{11}(ij) \left(2 J_{ij} - K_{ij}\right) + \nonumber \\
&&\sum_{i\ne j} {\mathsf s}(i) {\mathsf s}(j) p_{10}^{1/2}(ij)\,p_{01}^{1/2}(ij)\,
\xi(ij) \, K_{ij}, \label{energy}
\end{eqnarray}
where $J_{ij}$ and $K_{ij}$ are positive Hartree and exchange integrals
defined in terms of the NOs, 
$J_{ij} = \langle \phi_i\phi_i | \hat{w} | \phi_j\phi_j\rangle$ and 
$K_{ij} = \langle \phi_i\phi_j | \hat{w} | \phi_j\phi_i\rangle$.
With \rref{xi_DOCI} for $\xi(ij)$, $E$
in \rref{energy} is a functional of the NOs
and the 1- and 2-state OPs.
Kollmar introduced a similar J-K functional but simplified the 2-DM
\cite{Kollmar2004}.   
The signs are chosen by a sign rule and are not variables.
$p_{10}$ is
related to $p_{11}$  and $p_1$  by $p_1(i) = p_{11}(ij) +
p_{10}(ij)$ and is eliminated from the functional.
Each sum in the denominator of $\xi(ij)$ in \rref{xi_DOCI} is simplified 
by the sum rule of \rref{p11sum_DOCI} to, e.g.,
\begin{displaymath}
\sum_{k(\ne i,j)} p_{11}(ik) = \frac1{2} (N-2) p_1(i) - p_{11}(ij).
\end{displaymath}

As stated above, we assume that the $(2,2)$
and $(2,3)$ positivity conditions are sufficient in practice. Under
this circumstance, the infimum of $E$  with respect to the
NOs and the OPs, subject to the constraints \rref{rho1cond} and
\rref{bound1}, \rref{bound2}, \rref{p11sum_DOCI}, yields a variational
approximation to the ground-state energy, 
apart from the $\xi$-approximation \rref{xi_DOCI}, for $N>4$.
 
\rref{energy} is a generalization
of the NOFT formulations of 1-DM functional theories, which 
require only 1-state OPs. The extra complexity from
2-state OPs and implicit 4-state OPs is more than compensated by the substantial gain in
accuracy it makes possible. The computational cost of calculating $E$ 
from \rref{energy} scales like Hartree-Fock
energy-functional minimization with a greater prefactor ($M^3$ vs $(N/2)^3$)
due to fractional occupation of NOs. 

\section{Numerical results for simple molecular systems}

To test OP-NOFT-0, we studied several diatomic molecules,
the Paldus H$_4$ test, and
linear chains of H atoms with open boundary conditions.  We included
all electrons (core and valence) and expanded the NOs in the
Gaussian 6-31G$^{**}$, STO-6G, and cc-pVTZ  bases. The
constrained minimization was performed by damped
Car-Parrinello dynamics \cite{Car1985}, as detailed in {Section S2} of
the SI.
 
We started the minimization from NOs and OPs obeying the constraints
but otherwise random. The signs were taken from the sign rule of
the {Table S2} in {Sec.~S1} of the
SI. They were kept fixed during optimization.

At convergence, the {\em active} subset of NOs had $p_1 \ge
10^{-3}$. The remaining NOs contributed negligibly to the energy. The
same active NOs and signs were found for several test cases
starting instead from a sufficiently large set of random NOs, half with
positive and half with negative signs\footnote{For H$_2$ the sign rule
  (${\mathsf s}(i\le N)=+1$ and ${\mathsf s}(i>N)=-1$) holds near equilibrium, but a more
  complex pattern emerges at large separation where additional
  positive signs are needed for the van der Waals tail of the
  interaction potential \cite{Sheng2013}. In principle, these positive
  signs could be obtained with our minimization procedure, but their
  effect is beyond the accuracy of the present calculations}.  It is
significant for the rule of \rref{signproduct} for arbitrary
matrix-element signs that for the systems tested, the Brillouin-Wigner
perturbation-theory based rule of {Table S1}, and the
alternative of random initial 
assignment of signs to pairs yield the same results for positive
matrix elements. The procedure of \rref{signproduct} also yields half positive and half
negative signs for the pairs when signs are assigned randomly to the
individual NOs with no reference to the matrix-element signs.

We report the dissociation energy curves of the dimers H$_2$, LiH and HF
in {Figs.~S1, S2}, and {S3} in
{Section~S3} of the SI. 
We performed restricted Hartree-Fock, DFT (PBE \cite{Perdew1996} and/or
PBE0 \cite{Adamo1999}), CASSCF, and CCSD(T) calculations with the same
basis.
For H$_2$, our 
functional depends only on 1-state OPs and the signs ${\mathsf s}(i)$; it
reduces to the exact expression of Löwdin and Shull
\cite{Lowdin1956}. When the ${\mathsf s}(i)$ are chosen from {Table S2} of
the SI, the OP-NOFT-0 
dissociation energy curve coincides with that of CASSCF at all
interatomic separations, implying that this  sign rule is exact for
H$_2$. Even in a system as simple as H$_2$, spin-restricted Hartree-Fock and DFT
fail badly at dissociation because these single-reference theories
cannot recover the Heitler-London form of the wavefunction. 

OP-NOFT-0 becomes identical to DOCI for $N = 4$ when the sign choice is
correct. The conditions \rref{bound1},\rref{bound2}, and
\rref{p11sum_DOCI} simplify in this case as
discussed in {Section S4} of the SI.
Expression \rref{xi_DOCI} for $\xi$ is exact, but
the $A = 0$ restriction is not. We performed the H$_4$ Paldus test using a
minimal 1s basis set for OP-NOFT-0, DOCI, and FCI. The DOCI/OP-NOFT-0
equivalence and the accuracy of the OP-NOFT-0 signs are confirmed by
the results reported in {Section S5}. While DOCI only
captures about 25-90\% of the configuration-dependent correlation
energy in the Paldus test, the OP-NOFT-0 dissociation curve of LiH
almost coincides with CASSCF in {Fig.~S2}, which
indicates that its 1s-electrons are nearly inert so that higher
seniorities contribute negligibly to its ground-state energy.

HF, a 10-electron system, provides the first complete test of the
relation of OP-NOFT-0 to DOCI.  The energies obtained with the basis
set 6-31G$^{**}$ are shown in {Fig.~S3} of the SI. OP-NOFT-0 and DOCI are
above both CASSCF and CCSD(T) though OP-NOFT-0 lies below DOCI
because, while size consistent, the $\xi$-approximation is non-variational
here. The OP-NOFT-0 signs are correct. The results are discussed
further in {Section S3} of the SI.

\begin{figure}
\begin{center}
  \includegraphics[width= \columnwidth]{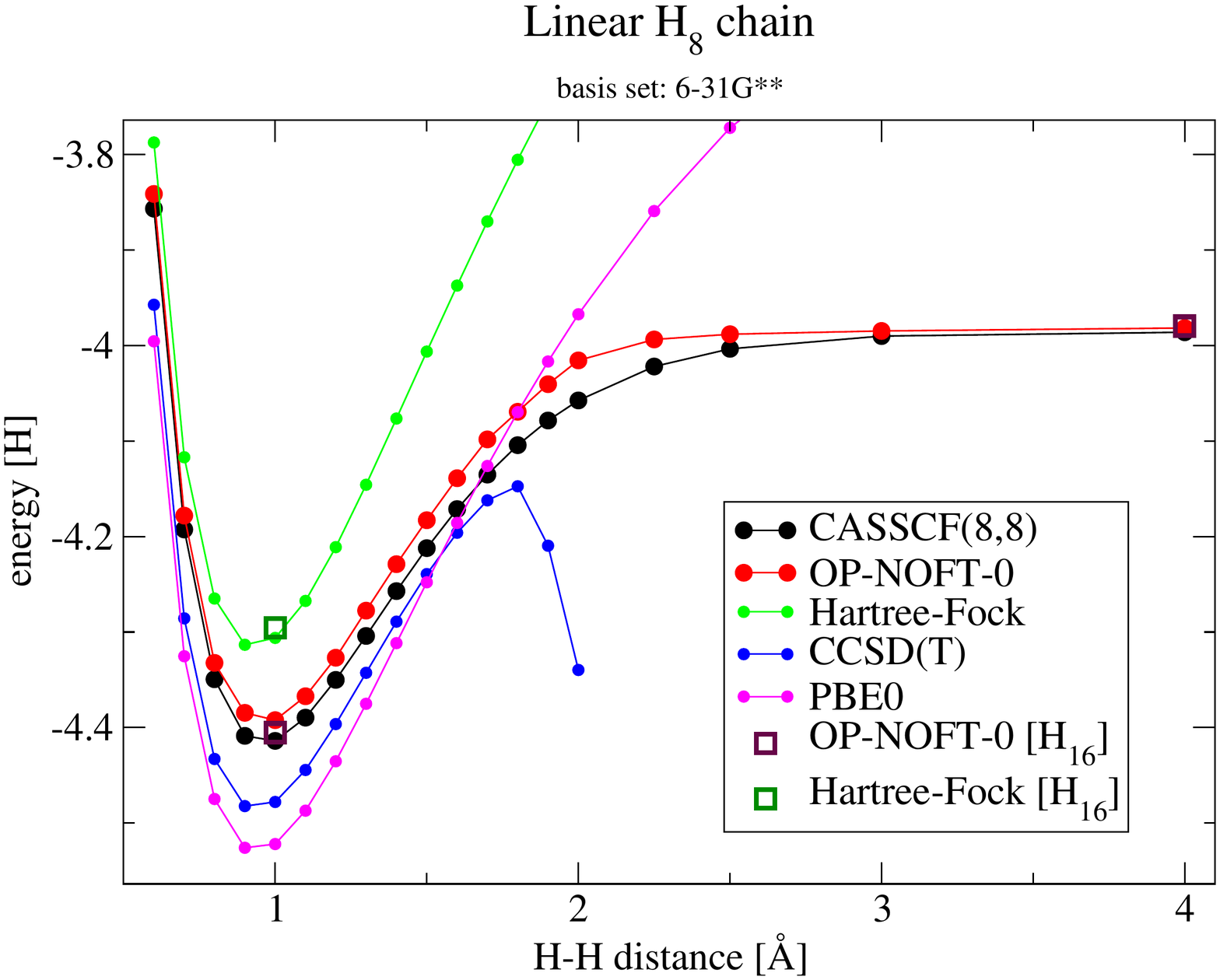}
  \caption{Symmetric dissociation curve of a linear H$_{8}$ chain.
    The squares indicate one half of the energy of a H$_{16}$ chain
    (black square: Hartree-Fock energy; green squares: OP-NOFT-0
    energy). All Hartree-Fock results are from spin-restricted
    calculations. \label{fig1}}
\end{center}
\end{figure}

Linear H chains are relatively simple systems
whose energy surfaces present a serious challenge for single reference
methods. Fig.~\ref{fig1} shows symmetric dissociation energy curves of
H$_8$. OP-NOFT-0 provides a consistent
description of the energy close to and everywhere above the CASSCF
reference. The breakdown of CCSD(T) at large separations is caused by
its single-reference character.  The deviation of
OP-NOFT-0 from CASSCF should be attributed mainly to the restriction
to the $A=0$ sector, a conclusion supported by the
seniority-restricted CI calculations of
Ref.~\cite{Bytautas2011}. Close comparison with those calculations is
not entirely straightforward, as Ref.~\cite{Bytautas2011} used the
slightly smaller 6-31G basis and a fixed, symmetric or broken
symmetry, molecular orbital (MO) basis, whereas we used
self-consistent NOs.  

The OP-NOFT-0 1-DM displays the entanglement due to correlation
through variation of the occupation numbers and the Von Neumann
entanglement entropy with interatomic separation shown in 
{Fig.~S5} in
{Section S6} of the SI. The increase of entanglement entropy with
separation signals a dramatic increase of correlation corresponding to
multi-reference character. The OP-NOFT-0 2-DM gives access to electron pair
correlations.

To test the dependence of the accuracy of OP-NOFT-0 on
electron number, we studied the symmetric dissociation of
H$_{16}$~\footnote{We do not give the CASSCF energies in this case, as
  the dimension of the active subspace would make these calculations
  very expensive.}.  Results for the energy of H$_{16}$ divided by 2
are shown as squares in Fig.~\ref{fig1}.  OP-NOFT-0 works equally well
for this longer chain.  The total energy at dissociation is twice that
of H$_{8}$, and the slightly increased binding energy per atom at
equilibrium arises from an increase in the correlation energy, as
expected from more effective screening in the larger system.

\begin{figure}
\begin{center}
  \includegraphics[width= \columnwidth]{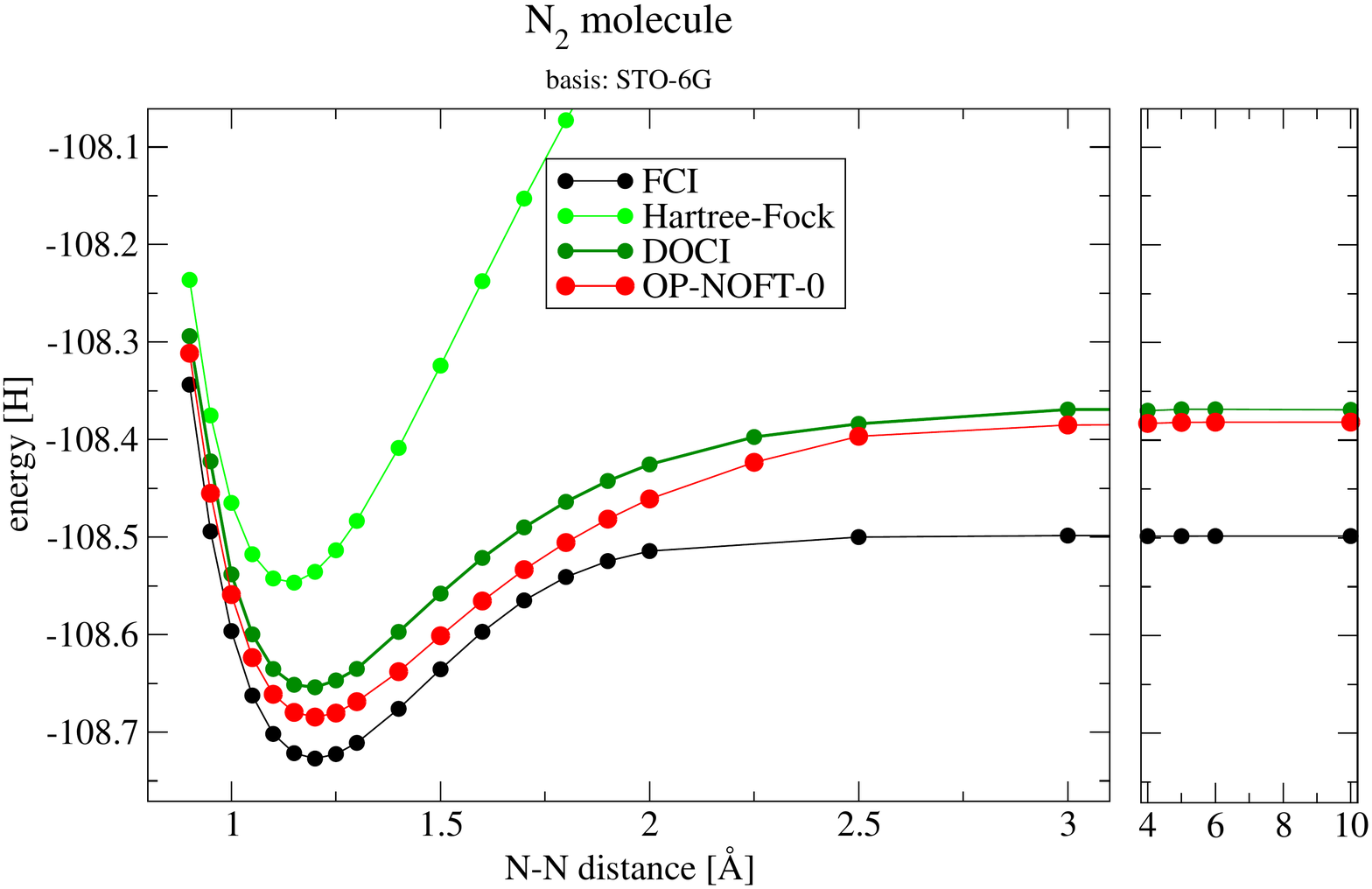}
  \caption{Dissociation curve of the N$_2$ molecule.
\label{fig2}}
\end{center}
\end{figure}

The N$_2$ molecule is a severe test because of its triple
bond. OP-NOFT-0, DOCI, and FCI results obtained with the minimal basis
set STO-6G are compared in Fig.~\ref{fig2}. They are similar to those for HF,
with OP-NOFT-0 and DOCI both above FCI but OP-NOFT-0 below DOCI. The
accuracy of our DOCI results was improved 
by use of the optimized OP-NOFT-0 NOs for the DOCI basis.

It is interesting to note that in all the systems studied, the
positivity condition $(2,2)$ was found to be sufficient at near
equilibrium up to intermediate separations dominated by dynamic
correlation because the $(2,3)$ condition was automatically satisfied
there. Moreover, only in the case of H$_8$, H$_{16}$ and N$_2$ at
large separations did inclusion of the $(2,3)$ positivity condition
turn out to be essential to prevent runaway from the ground-state
solution. 

\section{Discussion}

We have introduced a new method for correlated electronic-structure
calculations, OP-NOFT, that scales algebraically. Its DOCI-like
simplification, OP-NOFT-0, scales favorably with system size, with
Hartree-Fock 
energy-minimization scaling. The close correspondence of the energies
calculated via OP-NOFT-0 with DOCI
calculations support the accuracy of limiting the positivity
conditions to \rref{bound1}--\rref{bound2} and also of the
$\xi$-approximation \rref{xi_DOCI}.
OP-NOFT-0 is restricted  to the $A=0$ sector of 
the Hilbert space. It provides an accurate description of
single-bond breaking and is a considerable improvement over
single-reference methods in all cases studied.

Adding the computation of interatomic
forces to the OP-NOFT-0 energy-minimization methodology would make
possible the use of the theory for structural optimization and
ab-initio molecular dynamics~\cite{Car1985}.

From the practical point of view, minimization of the functional \rref{energy}
is significantly more laborious than minimization of the Hartree-Fock or the DFT
functional. We attribute this difficulty to the need
to include in \rref{energy} small occupation numbers. In
damped dynamics minimization the forces acting on the corresponding NOs
are thus very weak compared to the forces acting on the NOs with
occupation numbers close to 1, slowing down considerably the entire
procedure. This difficulty is common to all NO-based methods including
those based on the 1-DM. Solving it is essential to making OP-NOFT
methods applicable in practice.
Care must be taken to avoid spurious minima, as in other nonlinear
optimization problems.

\begin{acknowledgments}
The authors acknowledge illuminating discussions with Paul Ayers and
Kieron Burke.  Refs.~\cite{Weinhold1967} and~\cite{Weinhold1967a} were
brought to the authors' attention by Paul Ayers.  The authors further
wish to thank J.~E.~Moussa for important comments.
M.H.C. and R.C. acknowledge support from the DOE under grant
DE-FG02-05ER46201.
\end{acknowledgments}





\end{article}








\end{document}